\documentstyle[11pt, titlepage]{article}

\newcommand{\beq}{\begin{equation}}
\newcommand{\beqn}{\begin{eqnarray}}
\newcommand{\eeq}{\end{equation}}
\newcommand{\eeqn}{\end{eqnarray}}
\begin{document}
\begin{titlepage}
\rightline {HUTP-94/A033}
\rightline {DART-HEP-94/05}
\rightline {astro-ph/9408044}
\rightline {Forthcoming in {\it Phys. Rev. D}}
\begin{center}
\bigskip \bigskip \bigskip \bigskip \bigskip
\Large\bf Primordial Spectral Indices from \\
Generalized Einstein Theories \\
\bigskip\bigskip\bigskip
\normalsize\rm
David I. Kaiser \\
\bigskip \it Lyman Laboratory of Physics\\
Harvard University\\
Cambridge, MA 02138\\ \rm
e-mail:  dkaiser@fas.harvard.edu\\
\bigskip\bigskip\bigskip\bigskip
20 June 1995\\ \bigskip
[Revised Draft]\\ \bigskip
\bigskip\bigskip\bigskip\bigskip
\bf Abstract \rm
\end{center}
\narrower Primordial spectral indices are calculated to second order in
slow-roll parameters for three
closely-related models of inflation, all of which contain a scalar field
non-minimally coupled to the Ricci curvature scalar.  In most cases,
$n_s$ may be written as a function of the non-minimal curvature
coupling strength $\xi$ alone, with $n_s (\xi) \leq 1$, although the
constraints on $\xi$ differ greatly between \lq new inflation'
and \lq chaotic inflation' initial conditions.  Under \lq new inflation'
initial conditions, there are discrepancies between the values of $n_s$
as calculated in the Einstein frame and the Jordan frame.  The sources
for these discrepancies are addressed, and shown to have
negligible effects on the numerical predictions for $n_s$. No such
discrepancies affect the calculations under \lq chaotic inflation'
initial conditions. \bigskip\\

PACS numbers:  98.80C, 04.50 \\
\end{titlepage}
\newpage

\baselineskip 28pt
\section{Introduction}
\indent  In many models of the very early universe, the canonical
Einstein-Hilbert gravitational action emerges only as a low-energy effective
theory, rather than being assumed from the start.~\cite{adler}  A large
class of these generalized Einstein theories (GETs) involves scalar
fields non-minimally coupled to the Ricci curvature scalar.  Such
Brans-Dicke-like couplings~\cite{bd61} arise, for example, in models of
superstring compactification~\cite{string} and Kaluza-Klein
theories~\cite{freund}, and are related, via conformal transformation,
to quantum-gravitational counter-terms, which are proportional to the
square of the Ricci scalar.~\cite{hirai}~\cite{renorm} \\
\indent  Recent experimental determinations of the power spectrum of
density perturbations~\cite{exp}, modeled as $\cal{P} \rm \propto \it
k^{n_s} \rm$~\cite{physrep}, offer a rare glimpse of such Planck-scale
physics.  The spectral index for this scalar perturbation, $n_s$,
functions as a test for models of the very early universe, independently
of the familiar test based on the magnitude of perturbations.  It has
been shown, for example, that one well-known GET model of inflation,
extended inflation~\cite{EI}, cannot produce the observed nearly
scale-invariant (Harrison-Zel'dovich) spectrum:  extended inflation
predicts $n_s \leq 0.76$, instead of $n_s = 1.00$.~\cite{lidlyth92}  The
constraints on $n_s$ for extended inflation come from that model's
incorporation of a first order phase transition to exit inflation
(see~\cite{bubble} for more on this so-called \lq $\omega$ problem'.)
As discussed in~\cite{Spect}, this pitfall can be avoided in GET models
of inflation which undergo a second order phase transition to exit the
inflationary phase.  In this paper, three cousin-models of extended
inflation are considered, all of which fare much better in comparisons
with the observed values of $n_s$. \\
\indent  The analysis is carried out to second order in the
potential-slow-roll approximation (PSRA) parameters identified by Andrew
Liddle, Paul Parsons, and John Barrow~\cite{psra}, who have recently
amended earlier work by several authors~\cite{StewLyth}~\cite{relating}.
These papers are based on
the Hamilton-Jacobi equations of motion for a theory with a scalar
field minimally coupled to the curvature scalar; before they can be
applied to the non-minimally coupled GETs considered here, use must be
made of a conformal transformation~\cite{hirai}~\cite{conftrans}, which, via
field
redefinitions, puts the GET equations of motion into the \lq\lq Einstein
frame" form of an Einstein-Hilbert gravitational action with a minimally
coupled scalar field. \\
\indent  In this connection, it is important to keep
Redouane Fakir and Salman Habib's cautionary note in mind.
In~\cite{FakHab} they have demonstrated that ambiguities arise when
studying the quantum fluctuations of scalar fields in GETs in various
frames:
  the scalar two-point
correlation function evaluated in the \lq\lq Jordan" or \lq\lq physical"
frame, in which the non-minimal $\phi^2 R$ coupling is explicit, differs
from the two-point correlation function evaluated after the field
redefinitions, in the Einstein frame.  Yet, as discussed below, when the
inflationary expansion is
quasi-de Sitter, $a(t) \propto \exp(Ht)$, with $\dot{H} \simeq 0$, the
ambiguities isolated
in~\cite{FakHab} affect the {\it magnitude} of the correlation
function only, and not the $k$-dependence (and hence not $n_s$; see eq.
(58) in~\cite{FakHab}).  All three of the models
considered below display such quasi-de Sitter expansion under \lq\lq
chaotic inflation" initial conditions, and thus the Einstein frame
formalism employed here for $n_s$ should remain unproblematic.  \\
\indent  However, under \lq\lq new inflation" initial conditions,
two of the models evolve as a quasi-power-law, $a(t) \propto t^p$.
In these cases, ambiguities similar to those
discussed in~\cite{FakHab} {\it do} affect the form of $n_s$.  As discussed
below in section 3.2, the discrepancy between values of $n_s$ as
calculated in the Jordan and
Einstein frames arises because the curvature perturbation upon which the
PSRA formalism is based, ${\cal R} = (H / \dot{\phi}) \> \delta
\phi$~\cite{physrep}~\cite{StewLyth}, is not invariant with respect to
the conformal transformation.  (This discrepancy can be resolved by
choosing a suitable generalization of ${\cal R}$; see~\cite{DKpre}.)
Still, it can be shown that even in these cases of
quasi-power-law expansion, the numerical results for $n_s$ in the
Jordan frame, as calculated
with the PSRA formalism, differ negligibly from the Einstein frame
results. \\
\indent  The specific method for calculating $n_s$ is developed in
section 2.  In section 3, the formalism is applied to induced-gravity
inflation, for which we can compare the Einstein-frame results with
Jordan-frame calculations.  In sections 4 and 5, the analysis is
presented for two models with a different non-minimal $\phi^2 R$ coupling
and two different potentials.  Concluding remarks follow in section 6. \\
\section{Einstein frame formalism}
\indent  The calculation of $n_s$ for these GET models of inflation
involves two distinct tasks:  calculating the PSRA parameters, which
consist of various combinations of $d^n U / d\varphi^n$,
where $U$ is the scalar field potential following the conformal
transformation, and $\varphi$ is the newly-defined scalar field following
the conformal transformation; and calculating $\varphi_{HC}$, the value
of $\varphi$ when the scales of interest crossed outside of the horizon
during inflation.  In general the first of these tasks is
straightforward, while the second can become quite tricky. \\
\indent  The action for the three models studied below can be written in
the general form:
\beq
S = \int d^4 x \sqrt{-g} \left[ f(\phi) R - \frac{1}{2} \phi_{;\> \mu}
\phi^{;\> \mu} - V(\phi) \right] ,
\eeq
where $f(\phi)$ gives rise to the non-minimal coupling, $\phi^2 R$.  This
action yields the coupled field equations:
\beqn
\nonumber R_{\mu\nu} - \frac{1}{2} g_{\mu\nu} R &=& f^{-1}(\phi)
\left[\frac{1}{2} \left( \phi_{;\>\mu} \phi_{;\>\nu} -
\frac{1}{2} g_{\mu\nu} \phi_{; \>\lambda} \phi^{; \>\lambda} \right) +
f(\phi)_{;\>\mu\>;\>\nu} - \Box f(\phi) g_{\mu\nu}  -
\frac{1}{2} V(\phi) g_{\mu\nu} \right] , \\
\Box \phi &-& V^{\prime}(\phi) + f^{\prime}(\phi) f^{-1}(\phi) \left[3 \Box
f(\phi) +
\frac{1}{2} \phi_{;\>\lambda} \phi^{;\>\lambda} + 2V(\phi) \right] = 0 .
\eeqn
In eq. (2), a prime indicates $d / d\phi$. \\
\indent  These complicated field equations can be simplified by making a
particular conformal transformation (see, e.g.,~\cite{hirai}):
\beqn
\nonumber \hat{g}_{\mu\nu} &=& \Omega^2 (x) g_{\mu\nu} , \\
\Omega^2 (x) &=&  2 \kappa^2 f(\phi) ,
\eeqn
where quantities in the new frame are marked by a caret.  The quantity
$\kappa^2 = 8\pi M_{pl}^{-2}$, where $M_{pl} \simeq 1.22
\times 10^{19}$ GeV is the present value of the Planck mass.  (We thus
require that $f(\phi)$ remain positive definite, to ensure that $M_{pl}$
does not change sign.)  If we
further define a new scalar field $\varphi$ and scalar potential $U$ by:
\beqn
\nonumber \frac{d \varphi}{d \phi} &\equiv& \kappa^{-1} \sqrt{ \frac{f
(\phi) + 3 (f^{\prime}(\phi))^2}{2 f^2 (\phi)}} , \\
U(\varphi) &\equiv& (2 \kappa^2 f(\phi))^{-2}\> V(\phi) = \Omega^{-4}
\> V(\phi) ,
\eeqn
then the action in the new frame may be written in the canonical
Einstein-Hilbert form:
\beq
S = \int d^4 x \sqrt{-\hat{g}} \left[ \frac{1}{2 \kappa^2} \hat{R} -
\frac{1}{2} \varphi_{;\>\lambda} \varphi^{;\>\lambda} - U(\varphi)
\right] .
\eeq
The action in eq. (5) now yields the familiar equations of motion:
\beqn
\nonumber \hat{R}_{\mu\nu} - \frac{1}{2} \hat{g}_{\mu\nu}\hat{R} &=&
\kappa^2 \left[ \varphi_{;\>\mu} \varphi_{;\>\nu} - \frac{1}{2}
\hat{g}_{\mu\nu} \varphi_{;\>\lambda} \varphi^{;\>\lambda} - U(\varphi)
\hat{g}_{\mu\nu} \right] ,\\
\Box \varphi - U^{\prime}(\varphi) &=& 0 ,
\eeqn
where derivatives are now taken with respect to the metric
$\hat{g}_{\mu\nu}$, and a prime indicates $d / d\varphi$. \\
\indent  When evaluating the field equations, we will assume that the
background spacetime can be written in the form of a flat ($k = 0$)
Robertson-Walker line element:
\beqn
\nonumber ds^2 = g_{\mu\nu} dx^{\mu} dx^{\nu} &=& -dt^2 + a^2(t) d\vec{\rm
{x}}^2 \\
&=& \Omega^{-2}(x) \left( -d\hat{t}^2 + \hat{a}^2(\hat{t}) d\vec{\rm
{x}}^2 \right) ,
\eeqn
from which we can see that $d\hat{t} = \Omega (x) dt$, and $\hat{a}
(\hat{t} ) = \Omega (x) a(t)$.  These
relationships will become important when evaluating $\varphi_{HC}$. \\
\indent  The spectral index ($n_s$) is determined by~\cite{physrep}
\beq
n_s - 1 \equiv \frac{d \ln \delta_H^2}{d \ln k} ,
\eeq
where $\delta_H^2 = \vert \delta \tilde{\rho} / \rho \vert^2$.  An
exactly scale-invariant (Harrison-Zel'dovich) spectrum of perturbations
corresponds to $n_s = 1.00$.  For inflationary models, one can relate
deviations from this exactly scale-invariant spectrum directly to changes
in the Hubble parameter $\hat{H}(\varphi)$ and its derivatives during the
time that various scales were crossing outside of the
horizon.~\cite{StewLyth}~\cite{psra}~\cite{relating}  Such a
parametrization corresponds to the \lq\lq Hubble-slow-roll approximation"
(HSRA) scheme
of~\cite{psra}.  Using the inflationary equations of motion, these
deviations in terms of $\hat{H}(\varphi)$ can then be rewritten as
changes in
the inflaton's potential, $U(\varphi)$, and its derivatives.  This
parametrization corresponds to the \lq\lq Potential-slow-roll
approximation" (PSRA)
scheme of~\cite{psra}.  This is the approach adopted here.\\
\indent  To second order in PSRA parameters, the spectral index $n_s$
depends only on three parameters, $\epsilon$, $\eta$, and
$\zeta$.\footnote{To avoid confusion between the third PSRA parameter and
the non-minimal coupling strength, we will use $\zeta$ to denote the
PSRA parameter, and $\xi$ to denote the coupling strength.
In~\cite{psra}, the third PSRA parameter is labeled $\xi$, instead of
$\zeta$.}  These three functions of $\varphi$ are defined by~\cite{psra}:
\beqn
\nonumber  \epsilon &\equiv& \frac{1}{2 \kappa^2}
\left(\frac{U^{\prime}(\varphi)}{U(\varphi)} \right)^2 ,\\
\nonumber \eta &\equiv& \frac{1}{\kappa^2} \left(
\frac{U^{\prime\prime}(\varphi)}{U(\varphi)} \right) , \\
\zeta &\equiv& \frac{1}{\kappa^2} \left( \frac{U^{\prime}(\varphi)
U^{\prime\prime\prime}(\varphi)}{U^2(\varphi)} \right)^{1/2} ,
\eeqn
where, again, a prime denotes $d / d\varphi$.  To second order, then, the
spectral index is given by~\cite{StewLyth}~\cite{psra}:
\beq
n_s = 1 - 6\epsilon + 2\eta + \frac{1}{3} (44 - 18c) \epsilon^2 + (4c -
14) \epsilon \eta + \frac{2}{3} \eta^2 + \frac{1}{6} (13 - 3c) \zeta^2 ,
\eeq
where $c \equiv 4(\ln 2 + \gamma) \simeq 5.081$ ($\gamma \simeq 0.577$ is
Euler's constant). During inflation, each of these PSRA parameters
remains less than unity, and hence the deviations of the spectrum of
perturbations from the scale-invariant spectrum should indeed remain
small.\\
\indent  The PSRA parameters in eq.
(10) are to be evaluated at $\varphi_{HC}$.  Yet for two of the models
considered below, $\varphi(\phi)$ cannot be written in closed form.
Instead, the PSRA parameters can be written as functions of the
Jordan-frame scalar field $\phi$, by using eq. (4) and:
\beq
U^{\prime} = \frac{dU}{d\varphi} = \frac{d \phi}{d\varphi}
\frac{dU}{d\phi} ,
\eeq
and so on for the higher derivatives.  From eq. (4), it is clear that $U$
and all of its derivatives can always be written in closed form in terms
of $\phi$.  We can thus derive $\epsilon$, $\eta$, and $\zeta$ as
functions of $\phi$ alone.  This leaves the task of calculating the value
of $\phi$ which corresponds to $\varphi_{HC}$. \\
\indent  Solving for the value of the field at the time of
horizon-crossing is difficult in either frame; but,
following~\cite{dkprd}, we can use the fact that scales of interest to us
crossed outside of the horizon approximately 60 $e$-folds before the end of
inflation:
\beq
e^{\alpha} \equiv \frac{\hat{a}(\hat{t}_{end})}{\hat{a}(\hat{t}_{HC})} =
\frac{\Omega (x_{end})}{\Omega (x_{HC})}
\frac{a (t_{end})}{a (t_{HC})} \sim e^{60} .
\eeq
We can check how sensitively this assumption affects the calculation of
$n_s$; this is treated below, in section 3.1.  In
each of the models considered below, $a(t)$ can be solved in closed form
as a
function of $\phi (t)$ during the period of slow-roll, and since $\Omega
(x)$ is also defined as a function of $\phi (t)$, we may find an
approximate value for $\phi_{HC}$ in each case, where $\phi_{HC}$ is the
value of the Jordan-frame scalar field at the time when the scales of
interest crossed outside of the horizon {\it in the Einstein frame}. \rm
(See~\cite{guthjain} for a similar discussion in the context of extended
inflation.)  This last step allows the PSRA parameters to be evaluated at
the correct time. \\
\indent  Finally, it should be noted that by using the PSRA parameters
instead of the HSRA parameters, we have necessarily
made an additional assumption, referred to as the \lq\lq inflationary
attractor" assumption in~\cite{psra}.  That is, we have assumed that near
$\hat{t}_{HC}$, the full Einstein-frame field equations:
\beqn
\nonumber \hat{H}^2 + \frac{\hat{k}}{\hat{a}^2} &=& \frac{\kappa^2}{3}
\left[
U(\varphi) + \frac{1}{2} \left( \frac{d \varphi}{d \hat{t}} \right)^2
\right] , \\
\frac{d^2 \varphi}{d \hat{t}^2} + 3 \hat{H} \frac{d \varphi}{d \hat{t}}
&=& - U^{\prime}(\varphi) ,
\eeqn
may be approximated as:
\beqn
\nonumber \hat{H}^2 &\simeq& \frac{\kappa^2}{3} U(\varphi) , \\
3 \hat{H} \frac{d \varphi}{d \hat{t}} &\simeq& - U^{\prime}(\varphi) ,
\eeqn
where $\hat{H} \equiv \hat{a}^{-1} d \hat{a} / d\hat{t}$.  In other
words, we have assumed that the standard Einstein-frame \lq\lq slow-roll"
approximations may be made.  As discussed in~\cite{psra}~\cite{bondsal},
inflationary solutions of the full equations of motion, eq. (13),
approach the \lq\lq slow-roll" attractor situation, eq. (14), at least
exponentially quickly (provided that the sign of $d \varphi / d \hat{t}$,
and hence of $\dot{\phi}$, does not change -- we will assume this
here).  Thus, by using the PSRA parameters to study $n_s$, we assume
that $\hat{t}_{HC}$ occurs sufficiently late in the time-evolution of
the inflationary phase to allow the dynamics to converge on eq. (14).
It is this assumption which enables
the Jordan-frame scale factor $a(t)$ to be solved in terms of $\phi (t)$
during the slow-roll period. \\
\section{Induced-gravity Inflation}
\subsection{Einstein frame results}
\indent  The first model to be considered here is induced-gravity
inflation.~\cite{dkprd}~\cite{igi}  In this model, an extended
inflation-like Brans-Dicke coupling is combined with a Ginzburg-Landau
potential:
\beqn
\nonumber S &=& \int d^4 x \sqrt{-g} \left[ \frac{1}{2} \xi \phi^2 R -
\frac{1}{2} \phi_{;\>\mu} \phi^{;\>\mu} - V(\phi) \right] , \\
V(\phi) &=& \frac{\lambda}{4} \left( \phi^2 - v^2 \right)^2 ,
\eeqn
where $\xi \>\>(> 0)$ is the non-minimal coupling strength, and is related
to the
Brans-Dicke parameter $\omega$ by $\xi = (4 \omega)^{-1}$.  The
non-minimal coupling turns the Planck mass into a dynamical quantity; the
present value of the Planck mass is related to the vacuum expectation
value of the potential, $v$, by $M_{pl} = \sqrt{8 \pi \xi}\>v$.  In a flat
Friedmann universe, the Jordan-frame field equations are:
\beqn
\nonumber H^2 &=&  \frac{1}{3 \xi \phi^2} V(\phi) + \frac{1}{6 \xi} \left(
\frac{ \dot{\phi}}{\phi} \right)^2 - 2H \left( \frac{\dot{\phi}}{\phi}
\right) ,\\
\ddot{\phi} &+& 3H \dot{\phi} + \frac{\dot{\phi}^2}{\phi} = \frac{1}{(1 + 6
\xi)} \frac{1}{\phi} \left[ 4V(\phi) - \phi V^{\prime}(\phi)\right] ,
\eeqn
where overdots denote time derivatives, and primes denote $d/d\phi$; we
have assumed that the classical background field $\phi$ is sufficiently
homogenous, so that all spatial derivatives become negligible.  These
equations correspond exactly to the Einstein-frame equations (13).  The
Einstein-frame \lq\lq inflationary attractor" field equations (14) may
then be rewritten as:
\beqn
\nonumber \left( H + \frac{\dot{\phi}}{\phi} \right)^2 &\simeq&
\frac{1}{3 \xi \phi^2} V(\phi) , \\
3 \left( H \dot{\phi} + \frac{ \dot{\phi}^2}{\phi} \right) &\simeq&
\frac{1}{(1 +
6 \xi)} \frac{1}{\phi} \left[ 4 V(\phi) - \phi V^{\prime}(\phi) \right] .
\eeqn
Yet the assumption, $U(\varphi) \gg \frac{1}{2} (d \varphi / d \hat{t}
\>)^2$, is equivalent to $V(\phi) \gg \frac{1}{2} \dot{\phi}^2
(1 + 6\xi)$,
and thus it remains consistent further to simplify the field equations
during slow-roll as:
\beqn
\nonumber H^2 &\simeq& \frac{1}{3 \xi \phi^2} V(\phi) , \\
3H \dot{\phi} &\simeq& \frac{1}{(1 + 6\xi)} \frac{1}{\phi} \left[ 4
V(\phi) - \phi V^{\prime} (\phi) \right] .
\eeqn
These approximate equations may be integrated to yield the solutions:
\beqn
\nonumber \phi (t) &=& \phi_o \pm \sqrt{\frac{4 \lambda \xi}{3 (1 + 6
\xi)^2}} \> v^2 \> t , \\
\frac{a(t)}{a_B} &=& \left(\frac{\phi(t)}{\phi_o}\right)^{(1 + 6 \xi) /
4\xi} \exp \left[ \frac{(1 + 6 \xi)}{8 \xi v^2} \left( \phi_o^2 -
\phi^2 (t) \right) \right] .
\eeqn
In eq. (19), $\phi_o$ and $a_B$ are values at the beginning of the
inflationary epoch.  The $\pm$ in $\phi (t)$ is determined by the initial
conditions:  for a \lq\lq chaotic inflation" scenario, $\phi_o \gg v$, and
the $-$ should be used in the solution of $\phi (t)$; in a \lq\lq new
inflation" scenario, $\phi_o \ll v$, so the $+$ should be used in the
solution
for $\phi (t)$.  Thus we can see that with the chaotic inflation initial
condition, $a(t)$ is dominated by a quasi-de Sitter
expansion for early times ($a(t) \propto \exp(\phi_o \sqrt{\lambda / 3
\xi} \> t)$), whereas with the new inflation initial condition, $a(t)$ is
dominated by a quasi-power-law expansion at early times ($a(t) \propto
t^{(1 + 6\xi) / 4\xi}$). \\
\indent  We may now make the conformal transformation of eqs. (3-4), in
order to calculate the PSRA parameters.  The conformal factor $\Omega (x)$
for induced-gravity inflation is
simply proportional to the Jordan-frame field:  $\Omega (x) =
\sqrt{\kappa^2 \xi } \> \phi (t)$, and the new scalar field potential,
written as a function of the Jordan-frame field, becomes:
$U(\phi) = (\kappa^2 \xi \phi^2 )^{-2}\> V(\phi)$.  Finally, the
Einstein-frame scalar field is defined by:  $d\varphi / d\phi =
\sqrt{(1 + 6\xi) / \kappa^2 \xi \phi^2}$.  Using these
relationships, the PSRA parameters of eq. (9) become:
\beqn
\nonumber \epsilon &=& \frac{8 \xi}{(1 + 6 \xi)}\frac{v^4}{(\phi^2 -
v^2)^2} ,\\
\nonumber \eta &=& \frac{8\xi}{(1 + 6\xi)} \frac{v^2 \> (2v^2 - \phi^2
)}{(\phi^2 - v^2)^2} ,\\
\zeta &=& \frac{4 \sqrt{2} \xi}{(1 + 6 \xi)} \sqrt{\frac{ v^4 \> (\phi^2
- 4
v^2)}{(\phi^2 - v^2)^3}} .
\eeqn
Before we may evaluate $n_s$, we must calculate $\phi_{HC}$ using
eq. (12), for which we need $\phi_{end}$, the value of the
Jordan-frame field at the time inflation ends in the Einstein
frame.\footnote{Note that this
value of $\phi_{end}$ should be very close to the value of $\phi$ at the
time inflation ends in the Jordan frame, since as $\phi \rightarrow v$,
$\Omega(x) \rightarrow 1$, and the two frames coincide.}  Inflation ends
(in the Einstein frame) once $d^2 \hat{a} / d \hat{t}^2 = 0$ (instead of
being $< 0$).  To first order, this is determined by $\epsilon =
1$.~\cite{psra}
If we write $\phi_{end} = \beta (\xi) \> v$, then we may solve for $\beta$:
\beq
\beta \simeq \sqrt{\left| 1 \pm \sqrt{\frac{8\xi}{(1 + 6 \xi)}}\right|} ,
\eeq
where, again, the $\pm$ is determined by the initial conditions:  $+$ for
a chaotic inflation scenario, and $-$ for a new inflation scenario.  Note
that $8 \xi / (1 + 6 \xi) \leq 4/3$, so in both the chaotic and new
inflation scenarios, the end of inflation occurs close to $\phi = v$, as
expected. \\
\indent  If we next write $\phi_{HC} = m (\xi)\> v$, then equation (12)
becomes:
\beq
e^{\alpha} = \left( \frac{\beta}{m} \right)^{(1 + 10 \xi) / 4 \xi} \exp
\left[ \frac{(1 + 6 \xi)}{8 \xi} \left( m^2 - \beta^2 \right) \right] .
\eeq
In order to solve for $m (\xi)$ under chaotic inflation conditions, it is
helpful to rewrite eq. (22) as:
\beq
\frac{m}{\beta} = \exp \left[ \frac{(1 + 6\xi)}{2 (1 + 10\xi)} \left( m^2
- \beta^2 \right) - \frac{8 \xi\alpha}{2 (1 + 10 \xi)} \right]  .
\eeq
To remain consistent, $m / \beta \geq 1$ for the chaotic inflation
scenario, which requires
\beq
m_{ch} \geq \sqrt{ \beta^2 + \frac{8\xi\alpha}{(1 + 6\xi)}} ,
\eeq
where the subscript \lq\lq {\it ch}" is to remind us that this inequality is
to be satisfied under chaotic inflation conditions only. \\
\indent  For the new
inflation scenario, it is helpful to rewrite eq. (22) as:
\beq
\frac{4 \xi \alpha}{(1 + 6 \xi)} = - \frac{(1 + 10 \xi)}{(1 + 6 \xi)} \ln
\left(\frac{m}{\beta} \right) + \frac{1}{2} m^2 - \frac{1}{2} \beta^2 .
\eeq
As in~\cite{Spect}, this equation may now be solved approximately for $m$
under two limiting conditions: ($a$) $4\xi \alpha \ll (1 + 6 \xi)$, and
($b$) $4 \xi \alpha \gg (1 + 6 \xi)$.  However, as discussed below in
section 3.2, $\xi$ is strongly constrained in the new-inflation scenario
of this model, based on sufficient inflation requirements:  $\xi \leq 2.5
\times 10^{-3}$.  Thus we need only consider the case ($a$).  In
this limit, $m$ becomes:
\beq
m_{new} \simeq 1 - \sqrt{\frac{4 \xi \alpha}{(1 + 6\xi)} + \left(
\beta - 1 \right)^2} , \\
\eeq
where the subscript \lq\lq {\it new}" is to remind us that this
approximate solution for $m$ applies only under the new inflation
conditions.  Note that given the constraint on $\xi$, the $(\beta - 1)^2$
term will always remain over an order of magnitude smaller than the
$4\xi \alpha / (1 + 6\xi)$ term.\\
\indent  Using eqs. (20, 24, 26), the PSRA parameters may now be written
as functions of the non-minimal coupling strength $\xi$ alone:
\beqn
\nonumber \epsilon &=& \frac{8 \xi}{(1 + 6 \xi)} \frac{1}{(m^2 - 1)^2} ,\\
\nonumber \eta &=& \frac{8\xi}{(1 + 6 \xi)} \frac{(2 - m^2)}{(m^2 -1)^2} ,\\
\zeta &=& \frac{4 \sqrt{2} \xi}{(1 + 6 \xi)} \sqrt{\left|\frac{(m^2 -
4)}{(m^2 - 1)^3}\right|} ,
\eeqn
where the appropriate $m ( \xi )$ is determined by the initial conditions. \\
\indent  Approximate first-order results for $n_s$ may be written using
eq. (10), and taking the limits:  $m \gg 1$ for chaotic inflation initial
conditions, and $m \ll 1$ for new inflation initial conditions.
In these limits, to first order, the spectral index may be written:
\beqn
\nonumber n_{s,\>ch} &\simeq& 1 - \frac{16\xi}{(8\xi \alpha + 1)} , \\
n_{s, \> new} &\simeq& 1 - 16\xi ,
\eeqn
where we have used $\alpha \simeq 60 \gg 1$ when evaluating $n_{s,\>ch}$,
and $\xi \ll 1$ when evaluating $n_{s, \> new}$.
It is important to remember that these expressions for $n_s$ are limiting
cases, corresponding loosely to $\xi \geq \cal{O}\rm (1)$ and $\xi \leq
\cal{O}\rm (10^{-3})$, respectively;
the second-order result for $n_{s,\>ch}$, for example, has a
positive slope for increasing $\xi$, unlike this approximate solution.\\
\indent  The full second-order results for $n_s$ are plotted in Figure
1. For the chaotic inflation case, $0.90 \leq n_s \leq 0.97$ for
$\xi \geq 1.5 \times 10^{-3}$, which is obviously close to the observed $n_s
\sim 1.00$
spectrum~\cite{exp}.  For the new inflation case, $0.93 \leq n_s \leq 0.97$
for $\xi
\leq 4 \times 10^{-3}$. Thus, induced-gravity inflation predicts a
spectral index in close
agreement with the experimental determinations.  \\
\indent  The sensitivity of $n_s$ to our assumption $e^{\alpha} \simeq
e^{60}$ may be checked by calculating $\partial n_s / \partial \alpha$.
{}From eq. (10),
it is clear that this requires calculating $\partial \epsilon / \partial
\alpha$, $\partial
\eta / \partial \alpha$, and $\partial \zeta / \partial \alpha$.  For the
case of chaotic inflation conditions, eqs. (24) and (27) yield:
\beqn
\nonumber \frac{\partial \epsilon}{\partial \alpha}_{\vert_{ch}} &=&
\frac{-128 \xi^2}{(1 + 6\xi)^2} \frac{1}{(m^2 - 1)^3} , \\
\frac{\partial \eta}{\partial \alpha}_{\vert_{ch}} &=& \frac{64 \xi^2}{(1 +
6\xi)^2}\frac{(m^2 - 3)}{(m^2 - 1)^3} .
\eeqn
When $\xi \gg 1$, $m_{ch}^2 \rightarrow 4 \alpha / 3 \gg 1$, so
\beqn
\nonumber \frac{\partial \epsilon}{\partial \alpha}_{\vert_{ch}} &\propto&
\alpha^{-3} \sim \cal{O}\rm (10^{-6}) , \\
\frac {\partial \eta}{\partial \alpha}_{\vert_{\it ch}} \rm &\propto&
\alpha^{-2} \sim \cal{O}\rm (10^{-4}) .
\eeqn
When $\xi \ll 1$, $(m_{ch}^2 - 1) \rightarrow 8\xi \alpha / (1 + 6\xi)$,
so
\beqn
\nonumber \frac{\partial \epsilon}{\partial \alpha}_{\vert_{ch}} &\propto&
\xi^{-1} \alpha^{-3} , \\
\frac{\partial \eta}{\partial \alpha}_{\vert_{ch}} &\propto&
\xi^{-1} \alpha^{-3} \left(8\xi\alpha - 2\right) .
\eeqn
Because $\zeta$ only enters in $n_s$ at second order in the PSRA
expansion, $\partial \zeta / \partial \alpha$ has not been explicitly
included, although it can easily be shown to behave similarly to
$\partial \epsilon / \partial \alpha$ and $\partial \eta / \partial
\alpha$.  Likewise, for the new inflation case, eqs. (26) and (27) yield
$\partial \epsilon / \partial \alpha \propto \partial \eta / \partial
\epsilon \propto \xi^{4} \alpha^{2}$, so that both of these deviations
remain $\leq \cal{O}\rm (10^{-8})$, given the independent constraint for
new inflation conditions,
$\xi \leq 2.5 \times 10^{-3}$.  If we expand $n_s$ in a Taylor series as:
\beq
n_s (\alpha) \simeq n_s (60) + \left( \frac{\partial n_s}{\partial
\alpha} \right)_{\alpha = 60} \left( \alpha - 60 \right) + ... \> ,
\eeq
then eq. (31) can be used to place limits on the regions of
$\xi$-space, under chaotic inflation conditions, for which the assumption
$n_s (60)$ remains accurate:  requiring $\left( n_s
(\alpha) - n_s (60) \right) \leq 10^{-2}$, for $(\alpha - 60) \sim 10$,
limits $\xi$ to $\xi \geq 10^3 \alpha^{-3} \simeq \cal{O}\rm (10^{-3})$.
Note that under new inflation conditions, $\left( n_s (\alpha) - n_s (60)
\right)$ will always remain $\leq \cal{O}\rm (10^{-7})$.
  The assumption that $\alpha
= 60$, put in by hand to facilitate computation of $\phi_{HC}$, therefore
has negligible effects on the calculation of $n_s$, so long as $\xi \geq
\cal{O}\rm (10^{-3})$ under chaotic inflation conditions.  Thus, for the
remainder of this paper, all numerical results for $n_s$ will be
calculated assuming $\alpha = 60$. \\
\subsection{Comparison with Jordan frame results}
\indent  In~\cite{Spect}, $n_s$ was calculated directly in the Jordan
frame for a new inflation scenario of induced-gravity inflation.
There it was assumed that the scales of interest crossed outside of the
horizon while the expansion was still predominantly quasi-power-law,
$a(t) \propto t^p$, where $p = (1 + 6\xi) / 4\xi$.  The result was:
\beq
n_{s,\>J} = 1 - \frac{8\xi}{(1 + 2\xi)} .
\eeq
This should be compared with the $m \ll 1$ limit of the Einstein-frame
results in eq. (28):  $n_{s, \> E} \simeq 1 - 16\xi$.
Obviously, the results differ between the two frames. \\
\indent  The difference may be traced to ambiguities between the quantum
fluctuations of the scalar field in the two frames.~\cite{FakHab}  The
usual procedure for calculating the density perturbation spectrum is to
study the intrinsic curvature perturbation, given (during inflation)
by~\cite{physrep}~\cite{StewLyth}:
\beq
{\cal R} = \frac{H}{\dot{\phi}} \> \delta \phi .
\eeq
The spectrum of the density perturbation is then given by
\beq
{\cal P_R}^{1/2} = \frac{H}{\dot{\phi}} \left(  \vert \Delta \phi
\vert^2 \right)^{1/2} ,
\eeq
where $\vert \Delta \phi \vert^2$ is the two-point
correlation function for the scalar field's quantum fluctuations, defined
as~\cite{abwise}~\cite{guthpi}
\beq
\left| \Delta \phi (\vec{k}, \tau) \right|^2 \equiv k^3
\int \frac{d^3
x}{(2 \pi)^3} e^{i \vec{k} \cdot \vec{x}} \langle \delta \phi (\vec{x},
\tau) \delta \phi (\vec{0}, \tau) \rangle .
\eeq
This is the basis for the PSRA formalism.  The trouble is that although
${\cal R}$ is gauge-invariant with respect to the choice of comoving or
synchronous gauge~\cite{physrep}, it is {\it not} invariant with respect
to the conformal transformation of equations (3-4).  Labeling $\hat{\cal
R}$ the curvature perturbation as evaluated in the Einstein frame, it is
straightforward to show that
\beq
\hat{\cal R} \equiv \frac{\hat{H}}{d\varphi / d\hat{t}} \> \delta \varphi
= \frac{\left( H + \dot{\Omega}/\Omega \right)}{\sqrt{\dot{\phi}^2 + 6
\dot{\Omega}^2 / \kappa^2}} \> \Omega \> \delta \varphi \neq
\frac{H}{\dot{\phi}} \> \delta \phi ,
\eeq
or $\hat{\cal R} \neq {\cal R}$.  For induced-gravity inflation, $\Omega =
\sqrt{\kappa^2 \xi} \> \phi$, so
that during inflation, $\dot{\Omega} / \Omega = \dot{\phi} / \phi \ll
H$~\cite{FakUn}; similarly, under the new inflation conditions, with $\xi
\leq {\cal O} (10^{-3})$, then $6 \dot{\Omega}^2 / \kappa^2 = 6 \xi
\dot{\phi}^2 \ll
\dot{\phi}^2$, giving $\hat{\cal R} \simeq (H / \dot{\phi} ) \> \Omega \>
\delta \varphi$.  For
calculating $n_s$, however, it remains to compare the $k$-dependence of
$\delta \phi$ with that of $\delta \varphi$. \\
\indent  As
shown in~\cite{Spect}, near $t_{HC}$, the linearized equation of motion
for the fluctuations $\delta \phi$ is that of a nearly-massless scalar
field in an expanding background spacetime:
\beq
\ddot{\delta \phi} + 3H \dot{\delta \phi} - \frac{1}{a^2} \nabla^2 \delta
\phi \simeq 0 ,
\eeq
where $\ddot{\delta \phi} \equiv \partial^2 (\delta \phi) / \partial
t^2$.  Written in terms of conformal time, $d \tau \equiv dt /
a$, and a conformal field defined by $\psi \equiv a \delta \phi$, the
equation of motion for each mode becomes:
\beq
\tilde{\psi}_k^{\prime\prime} - \frac{a^{\prime\prime}}{a} \tilde{\psi}_k
+ k^2 \tilde{\psi}_k \simeq 0 ,
\eeq
where primes (in this section only) denote $d / d\tau$, and we have
performed a spatial Fourier transform.  For a metric expanding as $a(t)
\propto t^p$, the scale factor as a function of $\tau$ becomes $a(\tau)
\propto \left[ (1 - p) \tau \right]^{p/(1 - p)}$, and the equation of
motion takes the form:
\beq
\tilde{\psi}_k^{\prime\prime} + \left[ k^2 + \frac{p(1 - 2p)}{(1 - p)^2}
\frac{1}{\tau^2} \right] \tilde{\psi}_k \simeq 0 .
\eeq
Note that this approaches the equation of motion for a massless scalar
field in a de Sitter
background~\cite{FakHab}~\cite{abwise}~\cite{birdavies} as $p \rightarrow
\infty$, as it should given the form $a(t) \propto t^p$.  If we next
define the field $\chi$ as $\chi \equiv \tau^{-1/2} \psi$, and work in
terms of the variable $x \equiv k\tau$, then the equation of motion takes
the form of Bessel's equation:
\beq
\frac{d^2 \tilde{\chi}_k}{d x^2} + \frac{1}{x} \frac{d \tilde{\chi}_k}{d
x} + \left( 1 - \frac{1}{x^2} \left[ \frac{(3p - 1)^2}{4 (p - 1)^2}
\right] \right) \tilde{\chi}_k \simeq 0.
\eeq
Mode solutions for the original field $\delta \phi$ may then be written
in terms of Hankel functions:
\beq
\delta \tilde{\phi}_k (\tau) \simeq C_1 \tau^{\nu} \left[ A_k
H_{\nu}^{(1)} (k \tau) + B_k H_{\nu}^{(2)} (k \tau) \right] ,
\eeq
where $C_1$ is a constant, and
\beq
\nu = \nu_J = \frac{3p - 1}{2 (p - 1)} .
\eeq
The subscript \lq\lq $J$" is to remind us that this solution is for the
fluctuations in the Jordan frame.  Again we see asymptotic agreement with
the de Sitter case, which has $\nu_{deS} = 3/2$. \\
\indent  The Bunch-Davies vacuum, as defined for the case of de Sitter
expansion, corresponds to $A_k = 0$~\cite{FakHab}~\cite{bunch}; note that
such a choice of vacuum is warranted for the case of power-law expansion
as well, since, in the limit $p \rightarrow 0$, this vacuum choice yields
mode solutions which approach the ordinary Minkowski space solutions for
massless scalar particles, $\propto k^{-1/2} \exp (i \vec{k} \cdot \vec{x}
- ikt)$.~\cite{birdavies}~\cite{bunch2}  Taking the limit $k \tau
\rightarrow 0$ (for long-wavelength modes~\cite{abwise}), the
fluctuations $\delta \phi$ then behave, for this choice of vacuum, as
\beq
\delta \tilde{\phi}_k (\tau) \propto k^{-\nu_J} .
\eeq
The two-point correlation function for these fluctuations then becomes:
\beq
\left| \Delta \phi (\vec{k}, \tau) \right|_J^2 \propto k^{3 - 2 \nu_J} .
\eeq
This gives
\beq
{\cal P_R} \propto \delta_H^2
\propto \left| \Delta \phi \right|_J^2 \propto k^{3 - 2 \nu_J} .
\eeq
Using eqs. (8), (43), and $p = (1 + 6\xi) / 4 \xi$ yields the result:
\beq
n_{s,\> J} \simeq 1 - \frac{8\xi}{(1 + 2\xi)} .
\eeq
This is the origin of eq. (33). \\
\indent  The situation in the Einstein frame may now be compared:  the
fluctuations $\delta \varphi$ obey the equation of motion:
\beq
\frac{d^2 \delta \varphi}{d \hat{t}^2} + 3 \hat{H} \frac{d \delta
\varphi}{d \hat{t}} - \frac{1}{\hat{a}^2} \hat{\nabla}^2 \delta \varphi +
\frac{d^2 U (\varphi)}{d \varphi^2} \delta \varphi = 0 .
\eeq
At $\hat{t}_{HC}$, however,
\beq
\frac{\hat{k}^2}{\hat{a}^2 (\hat{t}_{HC})} = \hat{H}^2_{HC} \simeq
\frac{\kappa^2}{3} U(\varphi_{HC}) ,
\eeq
giving
\beq
\left( \frac{\hat{k}^2}{\hat{a}^2 (\hat{t}_{HC})} \right)^{-1} \frac{d^2 U
(\varphi)}{d \varphi^2} \propto \xi
\eeq
and under new inflation conditions $\xi \ll 1$, so in the Einstein frame
the
fluctuations $\delta \varphi$ also behave as a nearly-massless scalar
field. \\
\indent  The conformal transformation gives $\Omega \propto \phi \propto
t$, and thus $\hat{t} \propto t^2$.  Furthermore, $\hat{a} (\hat{t}) =
\Omega \> a(t)$, so a Jordan-frame scale factor $a(t) \propto t^p$
corresponds to an Einstein-frame scale factor $\hat{a} (\hat{t}) \propto
\hat{t}^{(p + 1) / 2}$.  The conformal transformation does not affect
$\vec{x}$, so $\hat{k} = k$.  Proceeding as above, the equation of motion
for the Einstein-frame fluctuations may be cast in the form of Bessel's
equation, and mode solutions written as:
\beq
\delta \tilde{\varphi}_k (\tau) = C_2 \tau^{\nu} \left[ \hat{A}_k
H_{\nu}^{(1)} (k \tau) + \hat{B}_k H_{\nu}^{(2)} (k \tau) \right] ,
\eeq
with $C_2 \neq C_1$ another constant, and
\beq
\nu = \nu_E = \frac{3p + 1}{2 ( p-1)} \neq \nu_J .
\eeq
Note that this result also yields the de Sitter solution, $\nu_{deS} =
3/2$, as $p \rightarrow \infty$. \\
\indent  If we attempt to define the Einstein-frame vacuum as $\hat{A}_k
= 0$, then the two-point correlation function for the fluctuations
$\delta \varphi$ becomes, as $k \tau \rightarrow 0$:
\beq
\left| \Delta \varphi (\vec{k} , \tau) \right|^2_E \propto k^{3 - 2
\nu_E} ,
\eeq
and thus
\beq
n_{s, \> E} = 1 - \frac{16\xi}{(1 + 2\xi)} .
\eeq
This reproduces the approximate first-order result for $n_{s, \> E}$, eq.
(28), given $\xi \leq \cal{O}\rm (10^{-3})$. \\
\indent  It is now easy to see why $n_s$ is unaffected by these
ambiguities for the case of quasi-de Sitter expansion:  when $a(t)
\propto \exp (Ht)$, the Jordan-frame two-point correlation function for
the quantum fluctuations
takes the asymptotic form~\cite{FakHab}~\cite{guthpi}:
\beq
\left| \Delta \phi (k \tau) \right|^2 \rightarrow C_3 \> \left[ 1 + (k
\tau)^2 \right] .
\eeq
We saw above that $k = \hat{k}$; similarly, the conformal time, $d\tau
\equiv dt / a$, remains invariant under the conformal transformation.
Thus, for quasi-de
Sitter inflation, the $k$-dependence does not change between the two
frames (although the magnitude of the correlation function does change
between the two frames~\cite{FakHab}), and the results for $n_s$ obtained
using the Einstein-frame formalism of section 2 should be accurate for
the Jordan frame as well. \\
\indent  Put another way, we may understand the discrepancy between the
two frames as follows:  quasi-de Sitter expansion in the Jordan frame
yields quasi-de Sitter expansion in the Einstein frame as well, so that
$\nu_J = \nu_E = 3/2$.  Quasi-power-law expansion in the Jordan frame
likewise gives quasi-power-law expansion in the Einstein frame, but with
a different power, so that $\nu_J \neq \nu_E$.  This difference in $\nu$
(if the vacua in the two frames are defined to be $A_k = \hat{A}_k = 0$)
is responsible for the different $k$-dependencies of $n_s$.  This
discrepancy may be remedied by finding a gauge-invariant measure of the
intrinsic curvature perturbation which {\it also} remains invariant with
respect to the conformal transformation.  (Note that the combination
presented in~\cite{FakUn} does not circumvent the discrepancy between
$\delta \phi$ and $\delta \varphi$ for models with quasi-power-law
expansion.)  Such a frame-independent formalism has been developed
in~\cite{DKpre}, with which the Jordan-frame value for $n_s$ does indeed
match the Einstein-frame PSRA result. \\
\indent  Finally, having calculated the discrepancy between $n_{s,\>J}$
and $n_{s,\>E}$ for the new inflation scenario of induced-gravity inflation,
it is important to consider how large a numerical
difference this ambiguity amounts to.  This can be done by finding the
allowed region of $\xi$-space which yields sufficient inflation.
In~\cite{psra}, Liddle, Parsons, and Barrow demonstrated that for
inflationary models to solve the horizon and flatness problems, the model
must provide at least 70 $e$-folds of expansion of the comoving Hubble
length, $(\hat{a} \hat{H})^{-1}$ (this is slightly different from the
requirement ordinarily assumed in the literature, that the scale factor
$\hat{a}$ grow by 70 $e$-folds).  To first order in the
PSRA parameters, this requires:
\beq
\bar{N}(\phi_o ,\phi_{end}) \equiv \ln
\frac{\left(\hat{a}\hat{H}\right)_{end}}{\left(\hat{a}\hat{H}\right)_{o}}
= - \sqrt{\frac{\kappa^2}{2}}
\int_{\phi_o}^{\phi_{end}} \frac{1}{\sqrt{\epsilon (\phi)}} \left( 1 -
\frac{1}{3} \epsilon (\phi) - \frac{1}{3} \eta (\phi) \right) d\phi \geq 70.
\eeq
Note that although $\bar{N}$ is written here in terms of the Jordan-frame
field $\phi$, it pertains, like the PSRA parameters $\epsilon$ and
$\eta$, to the Einstein frame; that is, we require that the comoving
Hubble length in the Einstein
frame inflate by at least 70 $e$-folds during inflation.  For the new
inflation scenario of induced-gravity inflation,
with $\phi_o \ll v$ and $\phi_{end} = \beta v$, this may be integrated
to yield the closed-form expression:
\beq
\bar{N} = \frac{2}{3 \sqrt{1 + 6 \xi}} \left( \beta - \ln \left(\frac{1 -
\beta}{1 + \beta} \right) \right) + \frac{\sqrt{1 + 6 \xi}}{4 \xi} \beta
\left( 1 - \frac{1}{3} \beta^2 \right) \geq 70 ,
\eeq
with $\beta (\xi)$ given in eq. (21).  This expression may be evaluated
numerically, revealing that $\bar{N} \geq 70$ for $\xi \leq 2.5 \times
10^{-3}$ (or $\bar{N} \geq 70$ for the Brans-Dicke parameter $\omega \geq
100$).\footnote{Note that for chaotic inflation conditions, additional
assumptions must be made about the value of $\phi_o$ before $\bar{N}$ can
be used to place limits on $\xi$.}  Considering the quasi-power-law
expansion for induced-gravity
inflation under the new inflation conditions, this result makes sense:
for small values of $\xi$, $a(t) \propto t^{1/ 4\xi}$, and thus $\xi \ll
1$ yields rapid expansion.  Furthermore, as discussed in~\cite{Spect},
there is no lower bound on $\xi$, as there is for extended inflation
(stemming from bubble percolation requirements), because of the second-order
phase transition in induced-gravity inflation.  This means
that the first-order result in the Jordan frame is $(1 - n_{s,\>J}) \leq
0.02$, while the first-order result in the Einstein frame is $(1 -
n_{s,\>E}) \leq 0.04$.  Thus, in either frame, induced-gravity inflation
with the new inflation conditions predicts $n_s \simeq 1$. \\
\indent  In the following two sections, the Einstein frame formalism of
section 2 is applied to two other closely-related GET models of
inflation. We only present the results for $n_s$ as determined by the
Einstein frame formalism of section 2; again, these should remain
invariant between the Einstein frame and the Jordan frame for the cases
of quasi-de Sitter inflation, and it is expected that the numerical
discrepancies between frames is small for the quasi-power-law case. \\
\section{Non-minimally coupled scalar with $\phi^4$ self-interaction}
\indent  The action, eq. (1), can be used to study a non-minimal
coupling similar to, but distinct from, that of induced-gravity
inflation.  In this section we consider a model given by the action (see,
e.g., ~\cite{FakUn}):
\beqn
\nonumber  S &=& \int d^4 x \sqrt{-g} \left[ \left(\frac{1 +  \kappa^2 \xi
\phi^2}{2
\kappa^2}\right) R - \frac{1}{2} \phi_{;\>\mu} \phi^{;\>\mu} - V(\phi)
\right] , \\
V(\phi) &=& \frac{\lambda}{4} \phi^4 ,
\eeqn
from which the Jordan-frame field equations in a flat Friedmann universe
become:
\beqn
\nonumber H^2 &=& \frac{\kappa^2}{3 (1 + \kappa^2 \xi \phi^2 )}
\left[ V(\phi) + \frac{1}{2} \dot{\phi}^2 - 6 \xi H \phi \dot{\phi}
\right] , \\
\nonumber  \ddot{\phi} + 3H \dot{\phi} &+& \left(\frac{\kappa^2 \xi \phi^2
(1 + 6\xi)}{1 + \kappa^2 \xi \phi^2 (1 + 6 \xi)} \right)
\frac{\dot{\phi}^2}{\phi} = \\
&=& \frac{1}{\left(1 + \kappa^2 \xi \phi^2 (1 +
6\xi)\right)} \left[ 4 \kappa^2 \xi \phi V(\phi) - (1 + \kappa^2 \xi
\phi^2 ) V^{\prime}(\phi) \right] ,
\eeqn
where overdots denote time derivatives and primes denote $d/d\phi$.  For
these field equations, the Einstein frame \lq\lq inflationary attractor"
assumption becomes:
\beqn
\nonumber H^2 &\simeq& \frac{\kappa^2}{3 (1 + \kappa^2 \xi\phi^2)}
V(\phi) , \\
3H \dot{\phi} &\simeq& \frac{1}{\left(1 + \kappa^2 \xi \phi^2 (1 +
6\xi)\right)} \left[ 4 \kappa^2 \xi \phi V(\phi) - (1 + \kappa^2 \xi
\phi^2) V^{\prime}(\phi) \right] .
\eeqn
As for induced-gravity inflation, these slow-roll field equations may be
integrated to yield $a(t)$:
\beq
\frac{a(t)}{a_B} = \left( \frac{1 + \kappa^2 \xi \phi^2 (t)}{1 + \kappa^2
\xi \phi_o^2} \right)^{3/4} \exp \left[ \frac{(1 + 6\xi)}{8} \kappa^2
\left( \phi_o^2 - \phi^2 (t) \right) \right] .
\eeq
Note from the form of the potential, $V(\phi)$, that this model only
admits chaotic inflation initial conditions, with $\phi_o \gg 0$, and
hence during inflation, the expansion is predominantly quasi-de Sitter. \\
\indent  A few words are in order concerning the sign of $\xi$ in this
model.
In induced-gravity inflation, the sign of $\xi$ is fixed by present
conditions: $\xi > 0$ is required to yield the proper value of the Planck
mass.  Yet
in this model, the sign of $\xi$ is undetermined by present conditions:
after inflation, $\phi \simeq 0$, and the present value of the Planck
mass is independent of the model parameters.  However, we
will only consider values of $\xi \geq 0$ here; as Toshifumi Futamase and
Kei-ichi Maeda concluded in~\cite{conftrans}, a negative value of $\xi$
(according to the sign conventions used here) would require $\left|
\xi\right|
\leq 10^{-3}$ in order to yield sufficient inflation.  Such
constraints do not apply for the sign choice $\xi \geq 0$.\\
\indent  Making the conformal transformation of eqs. (3-4) yields:
\beqn
\nonumber  \Omega (x) &=& \sqrt{1 + \kappa^2 \xi \phi^2 (t)} , \\
\nonumber  U (\phi) &=& \left(1 + \kappa^2 \xi \phi^2\right)^{-2} V(\phi)
, \\
\frac{d\varphi}{d\phi} &=& \frac{\left(1 + \kappa^2 \xi \phi^2 (1 + 6
\xi)\right)^{1/2}}{\left( 1 + \kappa^2 \xi \phi^2 \right)} .
\eeqn
The first of the PSRA parameters thus becomes:
\beq
\epsilon = \frac{8}{\kappa^2} \frac{1}{\phi^2 \left(1 + \kappa^2 \xi
\phi^2 (1 + 6 \xi)\right)} .
\eeq
Setting $\epsilon = 1$, and writing $\kappa^2 \xi \phi^2_{end} = \beta^2
(\xi)$, we may solve for $\beta (\xi)$:
\beq
\beta = \sqrt{\frac{1}{2 (1 + 6 \xi)} \left( \sqrt{192 \xi^2 + 32 \xi +
1} - 1 \right)} .
\eeq
Note that $\beta_{max} = 1.07$, and $\beta \rightarrow 0$ as $\xi
\rightarrow 0$. \\
\indent  If we similarly write $\kappa^2 \xi \phi^2_{HC} = m^2 (\xi)$,
then the three PSRA parameters may be written:
\beqn
\nonumber \epsilon &=& 8\xi\> \frac{1}{m^2 \left(1 + m^2 (1 +
6\xi)\right)} , \\
\nonumber \eta &=& 4\xi\> \frac{\left( 3 + m^2 (1 + 12\xi) - 2m^4 (1 +
6\xi)\right)}{m^2 \left( 1 + m^2 (1 + 6\xi) \right)^2} , \\
\zeta &=& 4 \sqrt{2} \xi\> \frac{\left| 3 + 2m^2 (-2 + 3\xi) - 15 m^4 (1 +
6\xi ) - 6m^6 (1 + 6\xi)^2 + 2m^8 (1 + 6\xi)^2 \right|^{1/2}}{m^2 \left( 1
+ m^2 (1 + 6\xi) \right)^2} .
\eeqn
As for induced-gravity inflation, $m(\xi)$ can now be approximated by
using eq. (12).  In this model, eq. (12) may be rewritten:
\beq
\frac{m^2}{\beta^2} = \frac{(1 + \beta^2)}{\beta^2} \exp \left[ \frac{(1
+ 6\xi)}{10 \xi} \left( m^2 - \beta^2 \right) - \frac{8\alpha}{10} \right]
- \frac{1}{\beta^2} ,
\eeq
where, again, the consistency of the chaotic inflation conditions
requires $m / \beta \geq 1$.  Remarkably, the requirements for this model
then take the same form as for induced-gravity inflation:
\beq
m_{ch} \geq \sqrt{ \beta^2 + \frac{8\xi\alpha}{(1 + 6\xi)}} ,
\eeq
with $\beta (\xi)$ for this model given by eq. (64). The spectral index
may now be calculated using eq. (10), where, again, the three
PSRA parameters of eq. (65) are functions of $\xi$ alone. \\
\indent  The first order result, $n_s = 1 - 6 \epsilon + 2 \eta$, may be
approximated in the limit $m \gg 1$, yielding:
\beq
n_s \simeq 1 - \frac{32\xi}{(16\xi \alpha - 1)} .
\eeq
Figure 2 shows the full second-order results.  The spectral index satisfies
$0.96 \leq n_s \leq
0.97$ for $\xi \geq 4 \times 10^{-3}$, again in very close agreement with
the empirical results. \\
\indent  The final model to be considered here is a close cousin to this
model, with the $\phi^4$ potential replaced by a Ginzburg-Landau
potential. \\
\section{Non-minimally coupled scalar with Ginzburg-Landau Potential}
\indent  As for the previous model, the action for this last model may be
written:
\beqn
\nonumber  S &=& \int d^4 x \sqrt{-g} \left[ \left( \frac{1 +  \kappa^2 \xi
\phi^2}{2
\kappa^2} \right) R - \frac{1}{2} \phi_{;\>\mu} \phi^{;\>\mu} - V(\phi)
\right] , \\
V(\phi) &=& \frac{\lambda}{4}\left(\phi^2 - v^2 \right)^2 .
\eeqn
There is a subtle difference between this model and the last, concerning
the present value of the Planck mass:  in this model, $\phi \simeq v$ at
the end of inflation, and not $\simeq 0$, which means that the present
value of the Planck mass is determined by:
\beq
M_{pl}^2 = \frac{8\pi \left( 1 + \kappa^2 \xi v^2\right)}{\kappa^2} .
\eeq
In other words, $\kappa^2 \neq 8\pi M_{pl}^{-2}$; instead, $\kappa^2$ is a
free
parameter of the model, with dimensions $mass^{-2}$.  For this reason, in
this section only, the present value of Newton's gravitational constant
will be written as: $\kappa_N^2 \equiv 8\pi M_{pl}^{-2} \neq \kappa^2$.
Equation (70) may then be rewritten:
\beq
\frac{\kappa^2}{\kappa_N^2} = 1 + \kappa^2 \xi v^2 \equiv 1 + \delta^2 ,
\eeq
where the parameter $\delta^2 \equiv \kappa^2 \xi v^2$ has been defined.
The spectral index in this model is thus a function of the two free
parameters, $\xi$ and $\delta^2$.
Note also that, as above, we will only consider values of $\xi \geq 0$. \\
\indent  The Jordan-frame slow-roll field equations, eq. (60), with the new
potential and $\kappa^2 \neq \kappa_N^2$, can then be integrated for $a(t)$:
\beq
\frac{a(t)}{a_B} = \left(\frac{1 + \kappa^2 \xi \phi^2(t)}{1 + \kappa^2
\xi \phi_o^2} \right)^{3/4} \left(\frac{\phi (t)}{\phi_o}
\right)^{\kappa_N^2 v^2 / 4} \exp \left[ \frac{(1 + 6\xi)}{8} \kappa_N^2
\left( \phi_o^2 - \phi^2 (t) \right) \right] .
\eeq
Although $\phi (t)$ cannot be solved exactly as a function of
$t$, the opposite can be done, to study how $\phi$ evolves with $t$.
That is, if we write the Jordan-frame field equations (60) as:
\beqn
\nonumber  H^2 &\simeq& \frac{\kappa^2}{3} \> A(\phi) \> V(\phi) , \\
3H \frac{d \phi}{dt} &\simeq& B(\phi) ,
\eeqn
then during slow-roll, we may invert this expression and integrate to
find $t$ as a function of $\phi$:
\beq
dt = \sqrt{3 \kappa^2} \> \frac{ \sqrt{A (\phi) \> V(\phi)}}{B(\phi)}\>
d\phi ,
\eeq
or:
\beqn
\nonumber t &=& \frac{ \pm 1}{\left(1 + \delta^2\right)}
\sqrt{\frac{3\kappa^2}{4
\lambda}} \left[ \ln \left( \frac{\phi (t)}{\phi_o} \frac{1 + \sqrt{1 +
\kappa^2 \xi \phi_o^2}}{1 + \sqrt{1 + \kappa^2 \xi \phi^2 (t)}} \right)
\right] \\
&\pm& \frac{(1+ 6\xi)}{\left(1 + \delta^2 \right)} \sqrt{\frac{3\kappa^2}{4
\lambda}} \left[ \left( \sqrt{1 + \kappa^2
\xi \phi^2 (t)} - \sqrt{1 + \kappa^2 \xi \phi_o^2} \right) \right] ,
\eeqn
where the $+$ is for new inflation initial conditions, and the $-$ for
chaotic inflation initial conditions.  It can be verified numerically
that the nonlinear terms in this expression are dominant only for very
small values of $\kappa^2 \xi \phi^2 (t)$, and that for $\kappa^2 \xi
\phi^2 (t) \sim \cal{O}\rm ( 1 )$, it is a good approximation to assume
$\phi (t) \propto t$.  Thus, as for induced-gravity inflation, the
expansion under chaotic inflation initial conditions is quasi-de Sitter,
while the expansion under new inflation initial conditions is
quasi-power-law. \\
\indent  The first PSRA parameter for this model is
\beq
\epsilon = \frac{8 (1 + \delta^2)}{\kappa_N^2} \frac{\phi^2}{\left(\phi^2 -
v^2 \right)^2} \frac{1}{\left(1 + \kappa^2 \xi \phi^2 (1 + 6\xi)\right)} .
\eeq
Writing $\kappa^2 \xi \phi_{end}^2 = \beta^2 (\xi)$, the equation
for $\beta$ becomes rather difficult to solve exactly:
\beqn
\nonumber 0 &=& \beta^6 (1 + 6\xi) + \beta^4 - 2\beta^3 \delta^2 (1 +
6\xi) \\
&+& \beta^2 \delta^2 \left( \delta^2 (1 + 6\xi) - \xi (1 + \delta^2)^2
\right) - 2\beta \delta^2 + \delta^4 .
\eeqn
Instead, for the chaotic inflation conditions, two limiting
approximations may be made: ($a$) $\phi_{end}^2
\gg v^2$ (i.e. $\beta^2 \gg \delta^2$), so that the $\beta$ calculated in
the previous section, eq. (64), can serve as the approximate value here; and
($b$) $\phi_{end}^2 \simeq
v^2$ ($\beta^2 \simeq \delta^2$).  We will also assume $\beta^2 \simeq
\delta^2$ for the new inflation initial conditions.  For case ($a$),
we will define $\kappa^2 \xi \phi^2_{HC} =
m^2 (\xi)$, and thus the PSRA parameters for case ($a$) may be written:
\beqn
\nonumber \epsilon_a &=& 8\xi (1 + \delta^2 )^2
\frac{m^2}{\left(m^2 - \delta^2 \right)^2 \left(1 + m^2 (1 + 6\xi)\right)}
, \\
\nonumber \eta_a &=& 4\xi (1 + \delta^2 ) \frac{\left(
F_a + G_a \delta^2 \right)}{\left(m^2 - \delta^2
\right)^2 \left(1 + m^2 (1 + 6\xi)\right)^2} , \\
\zeta_a &=& 4\sqrt{2} \xi (1 + \delta^2 )
\frac{\left| K_a + L_a \delta^2 \right|^{1/2}}{\left|m^2 - \delta^2
\right|^{3/2} \left(1 + m^2 (1 + 6\xi) \right)^2} ,
\eeqn
with $m^2 \gg \delta^2$, and
\beqn
\nonumber F_a &=& 3m^2 + m^4(1 + 12\xi) - 2m^6(1 + 6\xi) , \\
\nonumber G_a &=& -1 + 3m^2 + 4m^4(1 + 6\xi) , \\
\nonumber K_a &=& 3m^2 + 2m^4 (-2 + 3\xi) - 15 m^6 (1 + 6\xi) - 6 m^8 (1
+ 6\xi)^2 + 2m^{10}(1 + 6\xi)^2 , \\
L_a &=&m^2(7+12\xi) + 6m^4 (1 + 9\xi) - 9m^6(1 + 6\xi) - 8m^8 (1 +6\xi)^2 .
\eeqn
For both chaotic and new inflation conditions under case ($b$)
$\beta^2 \simeq \delta^2$, we will
define $\kappa^2 \xi \phi_{HC}^2 = m^2 (\xi) \> \delta^2$.  Then the
PSRA parameters become:
\beqn
\nonumber \epsilon_b &=& \frac{8\xi \left(1 + \delta^2 \right)^2}{\delta^2}
\frac{m^2}{\left( m^2 - 1 \right)^2 \left(1 + m^2 \delta^2 (1 + 6\xi)
\right)} , \\
\nonumber \eta_b &=& \frac{4\xi \left(1 + \delta^2 \right)}{\delta^2}
\frac{F_b}{\left(m^2 - 1 \right)^2 \left(1 + m^2 \delta^2 (1 +
6\xi)\right)^2} , \\
\zeta_b &=& \frac{4 \sqrt{2} \xi \left(1 + \delta^2\right)}{\delta^2}
\frac{ \left| K_b\right|^{1/2}}{\left|m^2 - 1 \right|^{3/2} \left(1 + m^2
\delta^2 (1 + 6\xi) \right)^2} ,
\eeqn
with
\beqn
\nonumber  F_b &=& -1 + 3m^2(1 + \delta^2) + m^4 \delta^2 \left( (1 +
12\xi) + 4\delta^2 (1 + 6\xi)\right) - 2m^6\delta^4 (1 + 6\xi) , \\
\nonumber  K_b &=& m^2 \left( 3 + \delta^2 (7 + 12\xi) \right) + 2m^4
\delta^2 \left( (-2 + 3\xi) + 3\delta^2(1 + 9\xi) \right) \\
&-& 3m^6 \delta^4
(1 + 6\xi) \left(5 + 3\delta^2 \right) - 2m^8\delta^6 (1 + 6\xi)^2
\left(3 + 4\delta^2\right) + 2m^{10}\delta^8(1 + 6\xi)^2 .
\eeqn
All that remains now is to calculate appropriate values for $m (\xi)$ for
each of these cases. \\
\indent  Proceeding as above, $m (\xi)$ becomes under each of these
conditions:
\beqn
\nonumber m_{ch,\> a} &\geq& \sqrt{\beta^2 + \frac{8\xi\alpha}{(1 + 6\xi)}
(1 + \delta^2)} , \\
\nonumber m_{ch,\> b} &\geq& \sqrt{1 + \frac{8\xi\alpha}{(1 + 6\xi)}
\frac{(1 + \delta^2)}{\delta^2}} , \\
m_{n,\> b} &\simeq& 1 - \sqrt{\frac{4\xi\alpha}{(1 + 6\xi)} \frac{(1 +
\delta^2)}{\delta^2}} ,
\eeqn
where $\beta$ for $m_{ch,\> a}$ is given by eq. (64).  Due to sufficient
inflation constraints in the new inflation case, we have only considered
$\xi \ll 1$.  The three first-order limiting cases for the spectral index
may thus be written:
\beqn
\nonumber n_{s,\>ch,\>a} &\simeq& 1 - \frac{32\xi}{(16\xi\alpha - 1)} , \\
\nonumber n_{s,\>ch,\>b} &\simeq& 1 - \frac{16\xi (1 +
\delta^2)}{( 8\xi\alpha (1 + \delta^2) + \delta^2 )} , \\
n_{s,\>n,\>b} &\simeq& 1 - 8\xi \frac{(1 + \delta^2)}{\delta^2} .
\eeqn
The full second-order results for various values of $\delta^2$ are shown
in Figures 3a and 3b.  As for the other models, $0.90 \leq n_s \leq
0.97$ for many
regions of allowed parameter space, yielding a spectral index in close
agreement with observed values. \\
\section{Conclusions}
\indent  The three closely-related GET models of inflation considered
above all predict values of $n_s$ close to the observed,
nearly-scale-invariant spectrum of perturbations.  For the
quasi-power-law cases (new inflation initial conditions), the
spectral index varies roughly linearly with the non-minimal
coupling constant $\xi$, with negative slope.  For large values of
$\xi$, then, this negative
slope-dependence of $n_s$ on $\xi$ could drag the predictions for $n_s$
below the experimentally observed values.  Yet
sufficient
inflation requirements place stringent restrictions on $\xi \ll 1$;
if such sufficient inflation requirements can be met, then
the resulting spectral index deviates only little from $n_s = 1.00$.  In the
quasi-de Sitter expansion cases (chaotic inflation initial conditions),
$n_s$ again varies roughly linearly with $\xi$, but with positive
slope; $n_s$ thus remains close to $n_s = 1.00$ for most values of $\xi$.
Note that these small deviations of $n_s$ from the
Harrison-Zel'dovich spectrum mean that each of the models considered
here predicts very small values for the tensor-mode perturbation index,
$n_T$, and the ratio of tensor to scalar mode amplitudes, $R$:  both
$n_T$ and $R$ are proportional to $\epsilon$ to first order~\cite{psra},
and in each of the cases above, $0 < \epsilon < \left|\eta\right| \ll 1$. \\
\indent  Under new inflation initial conditions, the Einstein frame
formalism employed here yields different
forms of $n_s (\xi)$ from calculations conducted exclusively in the
Jordan frame.  The physical basis for these discrepancies is
discussed in section 3.2, and is further treated in~\cite{DKpre}.  However,
again owing to the requirements from sufficient
inflation, in the allowed regions of $\xi$-space the numerical values
for $n_s$ differ negligibly between the two frames.  Under chaotic
inflation initial conditions, there are no discrepancies between the
forms of $n_s (\xi)$ in the two frames. \\
\indent  Each of these models is able to produce acceptable spectra, even
though their cousin-model extended inflation cannot, because they avoid
{\it both} of the so-called \lq $\omega$ problems' which plagued extended
inflation.~\cite{bubble}~\cite{lidlyth92}~\cite{Spect}  First, each of the
models considered here
exits inflation by slowly rolling towards the vacuum expectation value of
its potential, thereby avoiding the strict requirements from bubble
nucleation and percolation associated with a first order phase
transition.  This means that there is no lower bound on $\xi$ for these
models. Second, by exiting inflation, all three of these models {\it also}
exit the GET phase:  after inflation, as $\phi$ settles in to $v$ (or
$0$, for the $\lambda \phi^4$ model), the coefficient of the Ricci scalar
in the action, eq. (1), becomes the constant $1/(2\kappa_N^2)$.  Thus,
the second order phase transition responsible for ending inflation
simultaneously
delivers the universe into the canonical Einstein-Hilbert gravitational
form.  Unlike extended inflation, then, present-day tests of Brans-Dicke
gravitation versus Einsteinian general relativity place no restrictions
on allowed values of $\xi$ during the early universe. \\
\indent  The approach used in this paper can be generalized further, by
choosing a more general form for the GET action, eq. (1).  For example,
specifically \lq\lq stringy" effective actions, which often have the
\lq\lq wrong" sign for the kinetic term in eq. (1) and different
effective scalar potentials~\cite{string}, can be studied, as can models
with more than one scalar field coupled to the Ricci scalar
(e.g.,~\cite{hybrid}). By studying these GET models of inflation with
the methods employed here, we may further take advantage of the
window on Planck scale physics offered by the primordial spectrum of
density perturbations. \\
\section{Acknowledgments}
\indent  This research was conducted at Dartmouth College.  It is a
pleasure to thank Joseph Harris and Marcelo Gleiser for their
hospitality.  I would also like to thank Eric Carlson for many helpful
discussions, Andrew Liddle for his comments on an earlier draft of this
paper, Rich Haas for his help in
preparing the figures, and Redouane Fakir for supplying a copy of
ref.~\cite{FakHab}.  The very detailed comments from an anonymous
reviewer are also gratefully acknowledged.  This
research was supported by an NSF Fellowship for Predoctoral Fellows. \\
%


%

\newpage
\section{Figure Captions}

Figure 1.  Second-order results for the spectral index $n_s$
for induced-gravity inflation, based on equations (10, 24, 26, 27), with
$\alpha = 60$.  The solid line is for new inflation initial conditions,
and the dashed line is for chaotic inflation initial conditions. Note that
for the new inflation scenario, $\xi > 2.5 \times 10^{-3}$ is
forbidden, due to sufficient inflation requirements. \bigskip\\
Figure 2.  Second-order results for the spectral index $n_s$ for the
model of section 4, based on equations (10, 65, 67), with $\alpha = 60$.
This model only admits chaotic inflation initial conditions. \bigskip\\
Figure 3.  Second-order results for the spectral index $n_s$ for the
model of section 5, based on equations (10, 78-82), with $\alpha = 60$.
($a$)  Chaotic inflation initial conditions, with the
free parameter $\delta^2 = 10^{-6}$ (solid), $10^{-1}$ (dashed),
 $1$ (dot-dashed), and $10$ (dotted).
($b$) New inflation initial conditions, with the free
parameter $\delta^2 = 10^{-1}$ (solid), $1$ (dashed), and $10$
(dot-dashed).
\bigskip \\

\end{document}